\documentclass[aip,jcp,reprint]{revtex4-1}
\usepackage[applemac]{inputenc}
\usepackage{amssymb}
\usepackage{times}
\usepackage{graphicx}
\usepackage{subfigure}
\usepackage{natbib}
\usepackage{mciteplus}
\usepackage{dcolumn}
\usepackage{bm}

\begin{document}

\title{Revisiting Dynamics Near a Liquid-Liquid Phase Transition in Si and Ga: The Fragile-to-Strong Transition}

\author{Samuel Cajahuaringa}
\author{Maurice de Koning}
\email{dekoning@ifi.unicamp.br}
\author{Alex Antonelli}
\email{aantone@ifi.unicamp.br}
\affiliation{Instituto de F{\'i}sica Gleb Wataghin, Universidade Estadual de Campinas, UNICAMP, 13083-859 Campinas, S{\~a}o Paulo Brazil}

\begin{abstract}
Using molecular dynamics simulations we analyze the dynamics of two atomic liquids that display a liquid-liquid phase transition (LLPT): Si described by the Stillinger-Weber potential and Ga as modeled by the modified embedded-atom model (MEAM). In particular, our objective is to investigate the extent to which the presence of a dip in the self-intermediate scattering function is a manifestation of an excess of vibrational states at low frequencies and may be associated with a fragile-to-strong transition (FTST) across the LLPT, as suggested recently. Our results do not lend support to these suggestions. Specifically, in the case of Ga we observe the appearance of an excess of vibrational states at low frequencies, even in the absence of the appearance of a dip in the self-intermediate scattering function across the LLPT. Furthermore, studying the behavior of the shear viscosities traversing the LLPTs we find that, despite the development of a dip in the self-intermediate scattering function for the case of Si and its absence in Ga, both substances are fragile in character above and below their respective LLPT temperatures.
\end{abstract}

\date{\today}

\maketitle

\section{Introduction}
Liquid polyamorphism, i.e., the existence of distinct liquid phases of a pure substance, was proposed over 40 years ago~\cite{Rapoport1967} to explain the occurrence of unusual melting properties of certain substances, such as melting-curve maxima (as in P, Cs, Ba, etc) and negative-sloped melting lines (as in water, Si, Ge, Ga, etc).~\cite{McMillan2007} This so-called two-state model predicts the existence of a first-order phase transition between different liquid forms of the same substance, commonly referred to as the high-density liquid (HDL) and low-density liquid (LDL), with a coexistence line that ends at a second critical point. 

Although it is still an intensely debated issue,\cite{Barnes2009,Greaves2011,Wilding2013,Limmer2011,Liu2012} there are indications for the existence of such LLPTs in a number of substances. Aside from experimental observations of LLPTs in elemental phosphorus~\cite{Katayama2000}, $\rm Al_{2}O_{3}-Y_{2}O_{3}$,~\cite{Aasland1994,Greaves2008} the molecular liquids triphenyl phosphite~\cite{Tanaka2004,Kurita2004} and n-butane,~\cite{Kurita2005} and the compound AsS.~\cite{Brazhkin2008,Brazhkin2009} Other experimental data and computer simulations suggest the occurrence of LLPTs in other substances such as water,~\cite{Poole1992,Sciortino1997,Liu2005,Mallamace2008,Zhang2011} silicon,~\cite{Angell1999,Sastry2003,Miranda2004,Jakse2007,Ganesh2009,Garcez2011} gallium~\cite{Tien2006,Jara2009} and nitrogen.~\cite{Boates2009}

In addition to studies considering the structural characteristics associated with the transition between the HDL and LDL forms, there has been an increasing interest in the liquid dynamics across LLPTs.~\cite{Sastry2003,Poole2011,Cajahuaringa2012,Lad2012} In the case of silicon, Sastry and Angell~\cite{Sastry2003} found that, across an interval of 15~K through the LLPT, the intermediate scattering function develops a pronounced plateau characteristic of the so-called cage effect and the self-diffusion coefficient decreases by two orders of magnitude. A similar result was found in the case of gallium.~\cite{Cajahuaringa2012}

For silicon, the development of the pronounced caging plateau for LDL is accompanied by a pronounced dip in the intermediate scattering function just before the plateau.~\cite{Sastry2003} This dip has been linked to the appearance of an excess of low-frequency vibrational modes compared to the typical Debye model vibrational density of states (VDOS) of a crystal. This excess has been associated with a boson peak (BP),~\cite{Malinovsky1990,Nakayama2002,Binder2011} although this is still rather controversial.~\cite{Jakse2009,Yannopoulos2009} This observation, in turn, has led to the suggestion of a connection between the structural LLPT in Si and an accompanying transition in the dynamics between a fragile and strong liquid. Strong liquids are those whose transport properties, e.g., the shear viscosity, display an Arrhenius behavior as a function of temperature, whereas fragile liquids are characterized by a manifestly non-Arrhenius behavior. Empirically, the presence of a pronounced BP has been viewed as a signature of a liquid's strong character,~\cite{Sokolov1993,Rossler1996,Sokolov1997,Novikov2005} such as in the case of $\rm SiO_{2}$.~\cite{Binder2011,Horbach1996} In line with this perspective, the sudden appearance of a dip in the intermediate scattering function across the HDL-to-LDL transition in silicon, both for the Stillinger-Weber (SW) model~\cite{Stillinger1985} as well as in {\em ab initio} calculations, has been interpreted in terms of a fragile-to-strong transition (FTST).~\cite{Sastry2003,Jakse2008} 

This interpretation remains controversial, however,~\cite{Yannopoulos2000,Yannopoulos2009,Jakse2009} given that there is still no physical basis directly correlating the presence of a possible BP to liquid fragility. Indeed, as the fragility of a liquid is defined in terms of the temperature-dependence of transport properties, the verdict as to whether the appearance of an excess in the VDOS across the LLPT is accompanied by a FTST should be based on the analysis of transport properties such as the shear viscosity, which often appears in the so-called Angell plot~\cite{Angell1988} distinguishing between strong and fragile behavior.

Here we conduct a further investigation into the possible correlation between a change in dynamics and the occurrence of a LLPT. For this purpose we analyze the dynamics of liquid silicon and gallium as described by the SW potential and modified embedded-atom model (MEAM),~\cite{Baskes2002} respectively, both of which exhibit well-characterized LLPTs. For both systems we compute the intermediate scattering functions, the vibrational densities of states as well as the temperature-dependence of the shear viscosities. 

The remainder of the paper is organized as follows. Sec.~\ref{Sec1} provides the technical aspects of the computer simulations used in this study. The results are presented and discussed in Sec.~\ref{Sec2}. Finally, in Sec.~\ref{Sec3} we summarize the main results of our work.

\section{Computational Details}
\label{Sec1}

The computer simulations performed in this work were carried out using the molecular dynamics method (MD) as implemented in the LAMMPS package~\cite{Plimpton1995}. Isobaric-isothermal ensemble (NPT) simulations were employed to obtain the HDL and LDL phases, whereas canonical ensemble (NVT) simulations were used to compute correlation functions. The integration of the Nos\'{e}-Hoover equations of motion was carried out using a time step of 1 fs, and temperature and pressure damping parameters of 70 fs and 1 ps, respectively, were employed. The interatomic interactions in Si were modeled using the Stillinger-Weber (SW) \cite{Stillinger1985} potential, while Ga was described using the modified embedded atom method (MEAM) proposed by Baskes, Chen, and Cherne.~\cite{Baskes2002} The SW potential is known for providing an accurate description of liquid Si and its melting point temperature and it has been extensively used in the investigation of the LLPT in Si.~\cite{Angell1996,Sastry2003,Jakse2003,Vasisht2011} Although the MEAM potential for Ga overestimates the quantitative value of the melting point temperature at ambient pressure, it correctly reproduces essential qualitative features of the phase diagram such as the negative slope of the melting line of the $\alpha$-Ga phase, the positive slope of melting line of the GaII phase and the correct nine-fold coordination of the stable liquid. Moreover, the MEAM for Ga predicted a previously unknown crystalline metastable phase of Ga, which was confirmed by $\textit{ab initio}$ calculations.~\cite{deKoning2009} All calculations were based on simulation cells subject to periodic boundary conditions. In the cases of gallium and silicon, supercells containing 1152 and 1000 atoms were employed, respectively. Larger cells were used to evaluate finite-size effects on the dip in the intermediate scattering function in Si. 

Starting from the well-equilibrated liquid above the melting point temperature, the HDL and the LDL phases were obtained according to the following protocols. In the case of Si, the liquid was quenched starting at 1700~K at a cooling rate of 50~K/ns. For Ga, the quench started at 450~K at a cooling rate of 20~K/ns. In both cases the quenches were carried out at zero pressure, which is the same as that used in previous MD studies of these systems.~\cite{Sastry2003,Cajahuaringa2012}  The LLPT takes place around 1060~K in the case of Si,~\cite{Sastry2003} while for Ga it occurs at a temperature of 356~K.~\cite{Jara2009} The canonical-ensemble simulations of the HDL and LDL forms of Si were carried out at 1070~K and 1050~K, respectively. For the case of Ga the HDL and LDL forms were considered at 362~K and 350~K, respectively.

\section{Results and Discussion}
\label{Sec2}
As mentioned earlier, particular features in the intermediate scattering function, which describes the decay of density fluctuations on a given length scale,\cite{Hansen2006} have been interpreted as evidence for the occurrence of FTST in the case of the LLPT in Si. Here we compute the self- (or incoherent) intermediate scattering function, $F_S(\boldsymbol{k},t)$, of Si and Ga. For a system containing $N$ particles, this function is defined as
\begin{eqnarray}
\label{Fsk}
\nonumber
F_S(\boldsymbol{k},t)&=&\frac{1}{N} \sum_{i=1}^N \langle e^{i \boldsymbol{k}\cdot [\boldsymbol{r_i}(t)-\boldsymbol{r_i}(0)]}\rangle, \\
&=& \frac{1}{N} \sum_{i=1}^N F_{S,i}(\boldsymbol{k},t),
\end{eqnarray}
where $\boldsymbol{r}_{i}\left( t\right) $ stands for the position of particle $i$ at time $t$ and $\boldsymbol{k}$ is the wave vector associated with the length scale of interest. The magnitude of $\boldsymbol{k}$ is chosen to be $2\pi/a$, where $a$ corresponds to the position of the first peak in the radial distribution function. In the calculations for Si, we used the following values of $a$: 2.39~\AA, 2.43~\AA, 2.46~\AA, and 2.52~\AA, for the temperatures, 1050~K (LDL), 1070~K (HDL), 1200~K, and 1700~K, respectively. For Ga, the distances, 2.68~\AA, 2.66~\AA, 2.68~\AA, and 2.68~\AA, corresponding to the temperatures 350~K (LDL), 362~K (HDL), 400~K, and 450~K, respectively, were employed. All results reported below correspond to averages $F_S(k,t)$ of $F_S(\boldsymbol{k},t)$ over three mutually perpendicular $\boldsymbol{k}$-directions for a given magnitude $k$.

\begin{figure}[h!]
\includegraphics[width = 8.5 cm, angle=0 ]{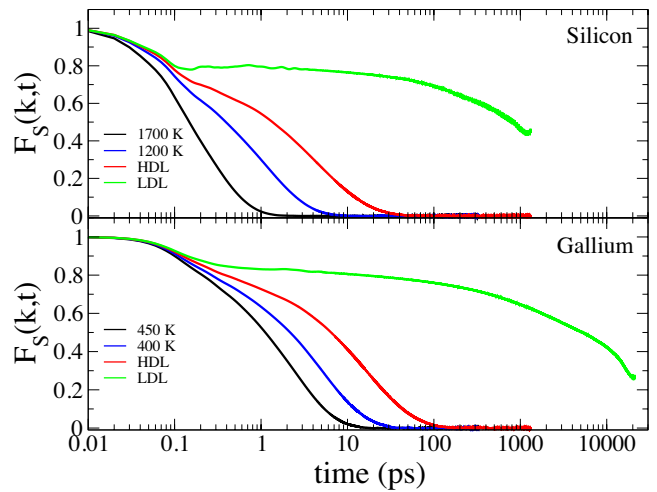}
\caption{Self-intermediate scattering function of liquid Si (upper panel) and liquid Ga (lower panel) for different temperatures. Upper panel: 1700~K, 1200~K, 1070~K (HDL) and 1050~K (LDL). Lower panel: 450~K, 400~K, 362~K (HDL) and 350~K (LDL).}
\label{Fig1}
\end{figure}

\begin{figure}[h!]
\includegraphics[width = 8.5 cm, angle=0 ]{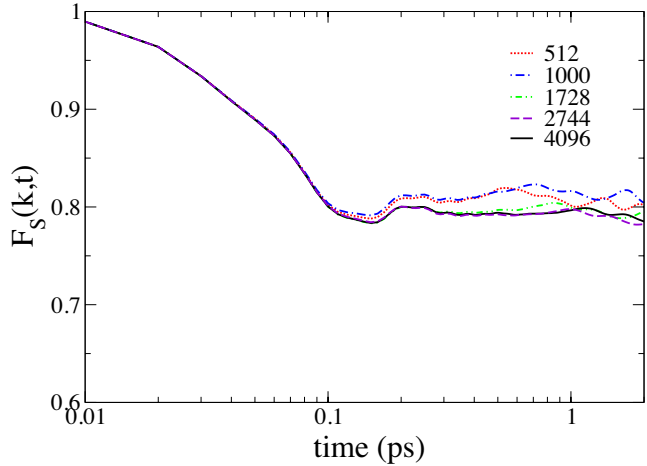}
\caption{Self-intermediate scattering function of LDL-Si for different simulation cell sizes: 512, 1000, 1728, 2744, and 4096 atoms.}
\label{Fig2}
\end{figure}

Fig.~\ref{Fig1} presents the results for $F_S(k,t)$ of Si and Ga for different temperatures above and below the LLPT.
Fig.~\ref{Fig1} displays the characteristic dip just at beginning of the plateau of $F_S(k,t)$ for LDL-Si, in agreement with previous studies.~\cite{Sastry2003,Jakse2007}  Yet, as pointed out by Horbach~\textit{et al.}~\cite{Horbach1996}, it is possible that the dip in the intermediate scattering function is an artifact of finite size effects. To verify this possibility we compute $F_S(k,t)$ for LDL-Si as a function of increasing cell sizes. The results are reported in Fig.~\ref{Fig2} and clearly display convergence for cell sizes above 2700 atoms, indicating that the appearance of the dip is indeed a robust effect. Returning to Fig.~\ref{Fig1}, in contrast to the case of Si, the LLPT in Ga is not accompanied by the appearance of a dip in $F_S(k,t)$.

According to the discussion in Ref.~\onlinecite{Jakse2008} this difference between the relaxation behaviors of LDL-Si and LDL-Ga, would indicate the presence of an excess in the vibrational density of states at lower frequencies with respect to the Debye model in the former and its absence in the latter. This would suggest significant differences between the LLPTs in Si and Ga from a dynamical point of view. To further investigate this issue, we compute the vibrational density of states (VDOS) of both LDL liquids, as determined by the Fourier transform of the normalized velocity autocorrelation function
\begin{equation}
\label{g(w)}
\nonumber
g\left( \omega\right) = \frac{1}{N}\sum_{j=1}^{N}\int_{-\infty}^{+\infty} \frac{\left\langle \boldsymbol{v}_{j}\left(  t\right)\cdot \boldsymbol{v}_{j}\left(0\right)\right\rangle }{\left\langle \boldsymbol{v}_{j}\left( 0 \right)\cdot \boldsymbol{v}_{j}\left(0\right)\right\rangle}e^{ i\omega t} dt,
\end{equation}
where $\boldsymbol{v}_{j}(t)$ is the velocity of the j-th particle at time $t$. 

Given that the VDOS described by the Debye model increases $\sim \omega^2$, the reduced VDOS, defined as, $g(\omega)/\omega^2$ (rVDOS) decays monotonically with increasing angular frequency. Therefore, an excess of vibrational modes at lower frequencies should appear in the form of peaks in the rVDOS. 

\begin{figure}[h!]
\includegraphics[width = 8.5 cm, angle=0 ]{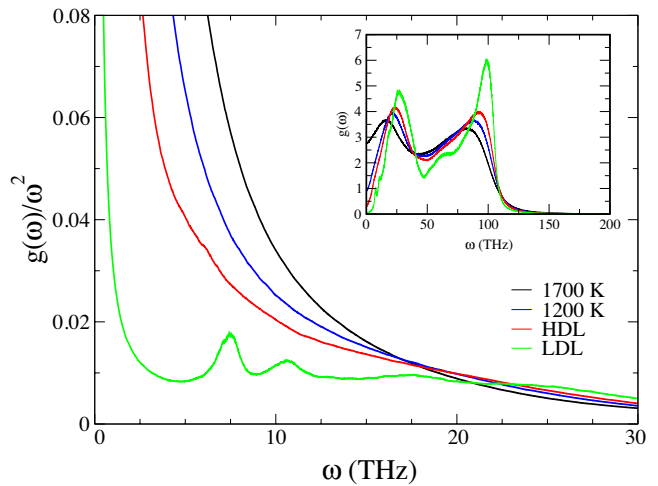}
\caption{Reduced VDOS and VDOS (inset) of Si as a function of angular frequency for different temperatures: 1700~K, 1200~K, 1070~K (HDL) and 1050~K (LDL).}
\label{Fig3}
\end{figure}

Fig.~\ref{Fig3} depicts the results of the rVDOS and VDOS for liquid Si at the same four temperatures shown in Fig.\ref{Fig1}. It is evident that for 1700~K, 1200~K, and 1070~K (HDL), the rVDOS decreases monotonically with increasing $\omega$. For the LDL liquid at 1050~K, on the other hand, it displays two distinct peaks at 7.5~THz and 11.0~THz, and a diffuse peak around 17.5~THz. This indicates that the LLPT in the case of Si, gives rise to an excess of low frequency modes, which could be associated with the BP. These results are in good agreement with those obtained by Jakse and Pasturel using \textit{ab initio} simulations.~\cite{Jakse2008} 

\begin{figure}[h!]
\includegraphics[width = 8.5 cm, angle=0 ]{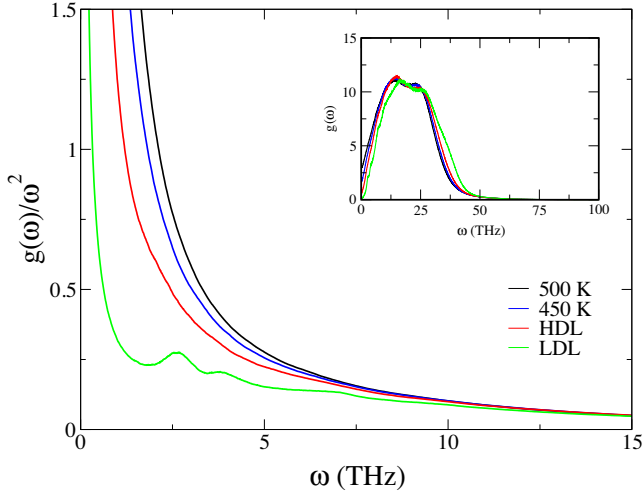}
\caption{rVDOS and VDOS (inset) of Ga as a function of angular frequency for different temperatures: 500~K, 450~K, 362~K (HDL) and 350~K (LDL)}
\label{Fig4}
\end{figure}

Fig.~\ref{Fig4} shows the results for the rVDOS and VDOS for liquid Ga at the same four temperatures considered in Fig\ref{Fig1}. Similar to the case of Si, for the temperatures 450~K, 400~K, and 362~K (HDL), the rVDOS decreases monotonically as $\omega$ increases. In addition, as for the LDL form of Si, the rVDOS of LDL-Ga at 350~K also exhibits two peaks between 2.5~THz and 4.0~THz, and a broad peak around 7.0~THz. This result is unexpected, however, if one assumes that an excess of vibrational modes at lower frequencies manifests itself as a dip at the beginning of the plateau in the intermediate scattering function $F_S(k,t)$, as has been suggested by Horbach and co-workers.~\cite{Horbach1996} In this view, the absence of a dip in $F_S(k,t)$ for LDL-Ga in Fig.~\ref{Fig1} should indicate the absence of an excess of vibrational modes at lower frequencies. Fig.~\ref{Fig4} shows that this is not the case for Ga, inconsistent with the proposed correlation between a dip in $F_S(k,t)$ and an excess of vibrational modes at low frequencies. 

Indeed, the identification of peaks in the rVDOS as evidence of the BP in the liquid is somewhat controversial. In the case of Si, for instance, there is experimental evidence that the BP is absent in amorphous Si,~\cite{Yannopoulos2009,Daisenberger2011} which appears at odds with the notion of the existence of such a peak in the precursor supercooled liquid phase.

Finally, we examine the suggested relationship between the appearance of dip in $F_S(k,t)$ traversing through a LLPT and a fragile-to-strong transition. For this purpose we compute the shear viscosity $\eta$ for both liquids as a function of temperature. The considered temperature ranges include conditions in which the liquids are thermodynamically stable, supercooled and below the LLPT temperature. The shear viscosity can be determined using a Green-Kubo-type relation involving the autocorrelation function of the off-diagonal component of the stress tensor,~\cite{Allen1989,Hansen2006}
\begin{equation}
\label{eta}
\nonumber
\eta = \frac{V}{k_{B}T}\int_{0}^{\infty}\langle\,  P_{\alpha \beta}(0) P_{\alpha \beta}(t) \,\rangle dt,
\end{equation}
where $V$ is the volume of the system, $T$ is the temperature, $k_{B}$ is the Boltzmann constant, and $P_{\alpha \beta}$ is an off-diagonal component of the stress tensor. Aside from the off-diagonal components $P_{xy}$, $P_{xz}$, and $P_{zy}$, there are, due to rotational invariance, two other independent components $\frac{1}{2}(P_{xx}-P_{yy})$ and $\frac{1}{2}(P_{yy}-P_{zz})$.~\cite{Alfe1998} In the calculations, $\eta$ is estimated using an average over these five independent components. 

\begin{figure}[h!]
\includegraphics[width = 8.5 cm, angle=0 ]{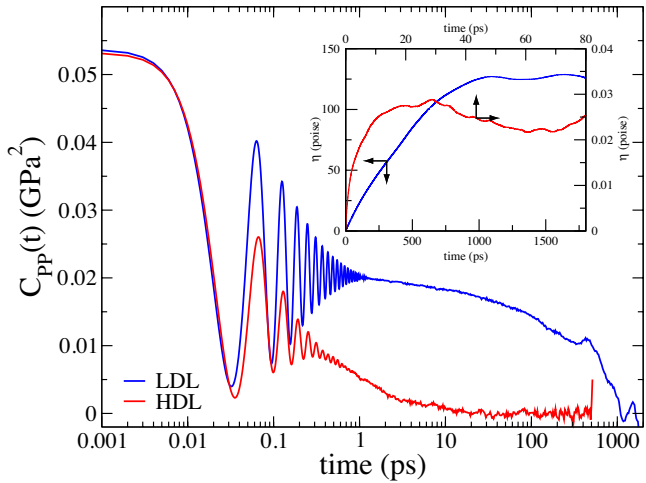}
\caption{Average of off-diagonal pressure-pressure autocorrelation function as a function of time for the HDL (red) and LDL (blue) forms of Si. Inset shows corresponding cumulative integrals for shear viscosities. Due to large scale differences, the HDL and LDL data are plotted with respect to different axis pairs, indicated by arrows.}
\label{Fig5}
\end{figure}

\begin{figure}[h!]
\includegraphics[width = 8.5 cm, angle=0 ]{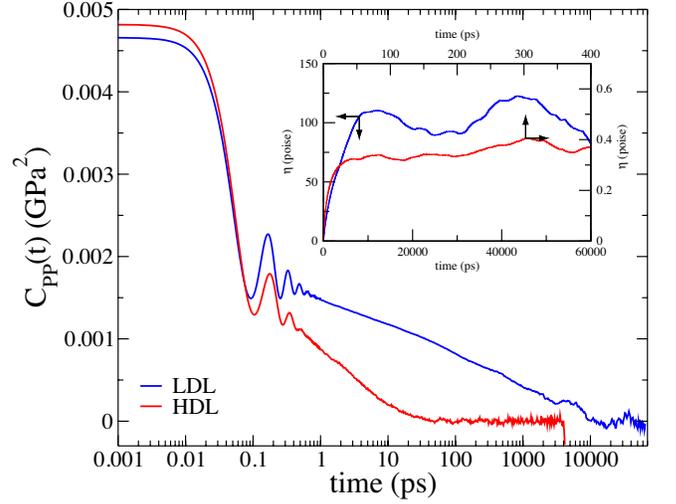}
\caption{Average of off-diagonal pressure-pressure autocorrelation function as a function of time for the HDL (red) and LDL (blue) forms of Ga. Inset shows corresponding cumulative integrals for shear viscosities. Due to large scale differences, the HDL and LDL data are plotted with respect to different axis pairs, indicated by arrows.}
\label{Fig6}
\end{figure}

Figs.~\ref{Fig5} and \ref{Fig6} show results for the average autocorrelation functions $C_{PP}(t)\,\equiv\, \langle \, P_{\alpha \beta}(0) P_{\alpha \beta}(t) \, \rangle$ and the corresponding cumulative integrals of Eq.~(\ref{eta}) for the HDL and LDL forms in Si and Ga, respectively. Both graphs display the convergence of the computed shear viscosities to plateau values and clearly show the dramatic scale changes occuring across the LLPTs.

\begin{figure}[h!]
\includegraphics[width = 8.5 cm, angle=0 ]{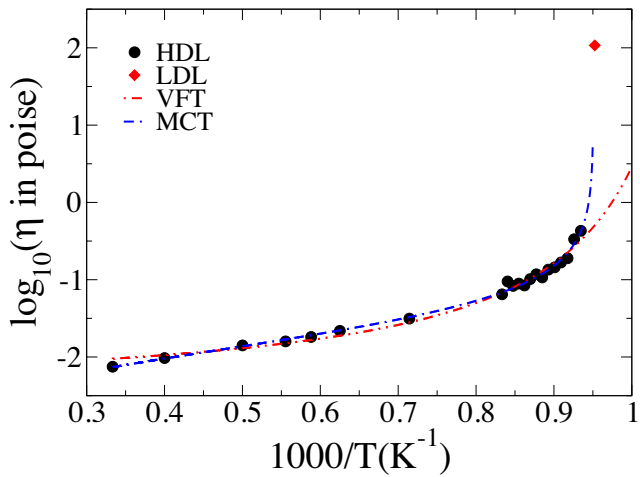}
\caption{Shear viscosity of liquid Si as a function of inverse temperature.}
\label{Fig7}
\end{figure}

The temperature-dependent results for liquid Si are depicted in Fig.\ref{Fig7}, showing the logarithm of the shear viscosity as a function of the inverse temperature. Starting at 3000~K, the obtained shear viscosity values, which are in agreement with experimental data,\cite{Rhim2000} increase steadily with decreasing temperature.  The increase is also manifestly nonlinear, with $\eta$ increasing more rapidly as the LLPT approaches. This point is further highlighted by the two curves obtained by fitting the temperature-dependence of the shear viscosity below the LLPT to the mode-coupling-theory (MCT) expression~\cite{Hansen1991,Hansen2006,Binder2011}
\begin{equation}
\label{mct}
\nonumber
\eta = \frac{\eta_{0}}{\left( T-T_{c}\right)^{\gamma}},
\end{equation}
where $T_{c}$ is the critical temperature at which the viscosity becomes singular, and the phenomenological Vogel-Fulcher-Tammann (VFT)~\cite{Vogel1921,Fulcher1925,Tammann1926,Binder2011} model
\begin{equation}
\label{vft}
\nonumber
\eta = \eta_{0} \exp\left( \frac{B}{T-T_{0}}\right),
\end{equation}
where the Vogel temperature $T_{0}$ describes the temperature at which the viscosity diverges. In the case of Si the least-squares regressions give $T_{c} = 1052$~K and $T_{0} = 877$~K, respectively, which affirm the rather strong non-Arrhenius behavior. This is consistent with the picture that Si is a fragile liquid for temperatures above LLPT. Upon passing through the first-order LLPT, $\eta$ undergoes a discontinuous increase, leading to a shear viscosity of that is two orders of magnitude larger than that of HDL-Si. This result is consistent with the calculations of Sastry and Angell~\cite{Sastry2003} that show an increase of two orders of magnitude of the diffusion coefficient upon transforming from the HDL to the LDL liquid forms. 

\begin{figure}[h!]
\includegraphics[width = 8.5 cm, angle=0 ]{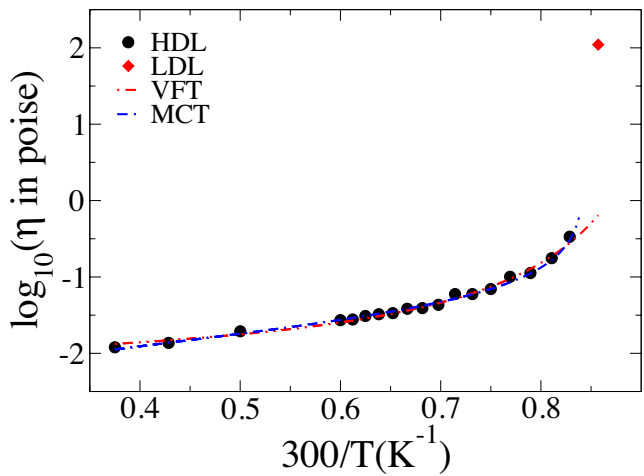}
\caption{Shear viscosity of liquid Ga as a function of inverse temperature.}
\label{Fig8}
\end{figure}

Fig.~\ref{Fig8} displays the same results for the case of Ga, including the curves obtained by fitting the data to the MCT and VFT expressions. The obtained values are consistent with available experimental data.~\cite{Spells1936}
Once more, Ga displays behavior that is similar to the case of Si. Above the LLPT the shear viscosity is manifestly non-Arrhenius, with critical temperature values of $T_{c} = 356$~K and $T_{0} = 301$~K, respectively. This implies that, as in the case of Si, Ga may be considered to be a fragile liquid for temperatures above the LLPT. Furthermore, the occurrence of the LLPT is also accompanied by an abrupt change of the viscosity, increasing about 2 orders of magnitude upon transforming from the HDL to the LDL. 

\begin{figure}[h!]
\includegraphics[width = 8.5 cm, angle=0 ]{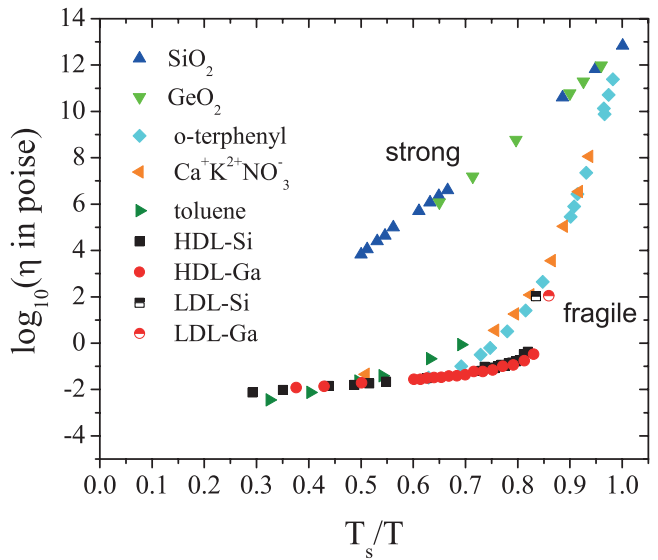}
\caption{Angell's plot. Our simulation results for shear viscosity of Si and Ga as a function of inverse temperature scaled by  $T_{s} = T_{0}$. For all other substances, the plot shows the experimental data for shear viscosity, extracted from Ref. \onlinecite{Angell1988}, as a function of inverse temperature scaled by $T_{s} = T_{g}$.}
\label{Fig9}
\end{figure}

To verify whether the abrupt increase of the shear viscosities of both LDL forms is indeed accompanied by a character change from fragile to strong, one should analyze the temperature dependences of both LDL shear viscosities. Unfortunately, such an explicit assessment is prohibitively costly in this case due to the extreme sluggishness of the dynamics of both LDL liquids on the time scale of the MD simulations. Therefore we adopt an indirect approach, comparing our LDL viscosity data for a single temperature (i.e., 1050~K and 350~K for Si and Ga, respectively) to those of typical fragile and strong liquids by plotting them in Angell's plot.~\cite{Angell1988} This plot depicts the logarithm of the viscosity as function of the inverse of temperature scaled by the material's glass-transition temperature $T_{g}$ and shows two distinct branches for strong and fragile liquids, respectively, in an essentially universal manner.~\cite{Angell1988} This is shown in Fig.~\ref{Fig9}, which displays a number of experimental viscosity data sets extracted from Ref.~\onlinecite{Angell1988} for various typical strong and fragile liquids, plotted as a function of $T_{s}/T$ where $T_s$ is the experimental glass-transition temperature $T_g$.

In principle, to plot the simulation results for Si and Ga on the same graph we need to determine the glass transition temperatures for the HDL and LDL phases of both substances. Unfortunately, however, an explicit determination of these $T_g$-values is unfeasible. For the HDL form it is obscured by the occurrence of the LLPT, whereas for the LDL it is inaccessible due to the extreme sluggishness of the dynamics. For this reason we need to infer the $T_g$'s by some indirect route. For the HDL phase we can establish a lower bound in terms of the Kauzmann temperature $T_K$ at which the excess entropy of the liquid with respect to the crystal vanishes.~\cite{Angell1997} Below $T_{K}$, the entropy of the liquid would be lower than that of the crystal, leading to the so-called entropy crisis, which is then avoided by the occurrence of the glass transition. In this view, $T_{K}$ can be regarded as a lower bound for $T_{g}$. In addition, it is well known that, for many substances,~\cite{Angell1997} the singularity temperature $T_{0}$ in the VFT equation is very close to $T_K$. We therefore use the divergence temperature $T_0$ of the VFT model determined from the HDL data in Figs.~\ref{Fig9} and ~\ref{Fig8} as estimates for $T_g$ of the HDL phases in Si and Ga.

Fig.~\ref{Fig9} shows the corresponding results for the shear viscosities of HDL Si and LDL Ga as a function of $T_{s}/T = T_{0}/T$. The results clearly show that both HDL phases of Si and Ga appear on the fragile branch of Angell's plot together with other fragile liquids such as toluene and o-terphenyl. Indeed, it is quite striking that the results for HDL Si and Ga, despite their very different structural and cohesive properties, almost superpose on the plot. 

An accurate placement of the LDL data on the plot is more challenging. Due to the extreme sluggishness of the dynamics below the LLPT it has proved prohibitive to obtain temperature-dependent data for the shear viscosity. For this reason we plot the LDL data using the divergence temperature $T_0$ obtained for the corresponding HDL phases to scale the inverse temperature. With this choice, despite the 2-order of magnitude increase of the shear viscosities across the LLPTs, the data points clearly remain in the fragile branch of the Angell plot. This suggests that, for both Si and Ga, the LLPT does not constitute a FTST. Evidently, this conclusion is based on the particular placement of the LDL results using the $T_0$ values obtained for the HDL phases as an indirect estimate for $T_g$ of the LDL forms. However, it seems plausible to assume that the HDL VFT divergence temperatures not only provide lower bounds for $T_g$ of the HDLs but also for the glass transition temperatures of the LDL phases. This is the case because the LDLs are much more viscous than their HDL counterparts, even for temperatures that are essentially the same. Thus it seems reasonable expect that the LDLs vitrify at higher temperatures than their HDL counterparts. In this line of reasoning, the representation of the LDL results in terms of the actual glass transition temperatures $T_g$ instead of $T_0$ would lead to a shift to the right in Fig.~\ref{Fig9}. Accordingly, the availability of the correct value of $T_g$ would still lead to the same conclusion reached based on the use of the VFT divergence temperatures $T_0$: despite the 2-order of magnitude increase of the shear viscosities across the LLPTs, the LDL phases still appear to belong to the fragile branch of Angell's plot. 

These findings are inconsistent with the occurrence of a FTST associated with the LLPTs in both Si and Ga. In fact, it appears that the LLPTs are accompanied by a transition from a fragile liquid to a less fragile liquid, regardless of the appearance or not of a dip in the intermediate scattering function. This seems at odds with the suggestion that the appearance of a dip in the intermediate scattering function through the LLPT signals a change of character from fragile to strong.~\cite{Sastry2003,Jakse2008,Yannopoulos2009,Jakse2009}

\section{Conclusions}
\label{Sec3}
We have performed a series of MD simulations to study the dynamic properties of supercooled liquid Si and Ga with the goal of investigating correlations between liquid dynamics and the presence of a structural LLPT.  The results indicate that, consistent with previous calculations, the intermediate scattering function of the LDL phase of Si at 1050~K exhibits a dip at the beginning of the plateau characteristic of the $\beta$-relaxation process which is absent for the HDL phase at 1070~K. For the case of the LLPT in Ga, on the other hand, the transition between the HDL to LDL forms is not accompanied by the development of a dip in the intermediate scattering function. 

Previous work~\cite{Sastry2003,Jakse2008} has suggested that the appearance of a dip in $F_S(k,t)$ across the LLPT is a manifestation of the development of an excess of vibrational states at low frequencies. In the context of our results for $F_S(k,t)$ this would imply a distinct difference between the dynamic behaviors of LDL Si and Ga, with the former characterized by an excess of low-frequency modes that should be missing for the latter. Our rVDOS results are at odds with this picture, however. Despite the absence of a dip in $F_S(k,t)$ for LDL-Ga, its rVDOS does display peaks characteristic of an excess of low-frequency modes.

In addition, we also consider the purported connection between the development of the dip in $F_S(k,t)$ and a transition in the character of liquid dynamics from fragile to strong. Within this view, the results obtained for $F_S(k,t)$ should point at a FTST accompanying the HDL-LDL structural transition for the case of Si and the absence of such a changeover in the case of Ga. Although the $T_g$ values for the LDL forms were estimated in an indirect manner, our results for the temperature dependence of the shear viscosities for both substances do not corroborate this prediction, indicating a view in which the HDL as well as LDL phases of both substances are fragile in character. In fact, it appears that the structural LLPT is accompanied by a transition from a fragile liquid to a less fragile liquid, regardless of the appearance or not of a dip in the intermediate scattering function.

\section*{Acknowledgments}

We gratefully acknowledge support from the Brazilian agencies CNPq, Fapesp and Capes. Part of the calculations were performed at CCJDR-IFGW-UNICAMP and CENAPAD-SP.

\bibliographystyle{apsrev4-1}

\begin{thebibliography}{62}%
\makeatletter
\providecommand \@ifxundefined [1]{%
 \@ifx{#1\undefined}
}%
\providecommand \@ifnum [1]{%
 \ifnum #1\expandafter \@firstoftwo
 \else \expandafter \@secondoftwo
 \fi
}%
\providecommand \@ifx [1]{%
 \ifx #1\expandafter \@firstoftwo
 \else \expandafter \@secondoftwo
 \fi
}%
\providecommand \natexlab [1]{#1}%
\providecommand \enquote  [1]{``#1''}%
\providecommand \bibnamefont  [1]{#1}%
\providecommand \bibfnamefont [1]{#1}%
\providecommand \citenamefont [1]{#1}%
\providecommand \href@noop [0]{\@secondoftwo}%
\providecommand \href [0]{\begingroup \@sanitize@url \@href}%
\providecommand \@href[1]{\@@startlink{#1}\@@href}%
\providecommand \@@href[1]{\endgroup#1\@@endlink}%
\providecommand \@sanitize@url [0]{\catcode `\\12\catcode `\$12\catcode
  `\&12\catcode `\#12\catcode `\^12\catcode `\_12\catcode `\%12\relax}%
\providecommand \@@startlink[1]{}%
\providecommand \@@endlink[0]{}%
\providecommand \url  [0]{\begingroup\@sanitize@url \@url }%
\providecommand \@url [1]{\endgroup\@href {#1}{\urlprefix }}%
\providecommand \urlprefix  [0]{URL }%
\providecommand \Eprint [0]{\href }%
\providecommand \doibase [0]{http://dx.doi.org/}%
\providecommand \selectlanguage [0]{\@gobble}%
\providecommand \bibinfo  [0]{\@secondoftwo}%
\providecommand \bibfield  [0]{\@secondoftwo}%
\providecommand \translation [1]{[#1]}%
\providecommand \BibitemOpen [0]{}%
\providecommand \bibitemStop [0]{}%
\providecommand \bibitemNoStop [0]{.\EOS\space}%
\providecommand \EOS [0]{\spacefactor3000\relax}%
\providecommand \BibitemShut  [1]{\csname bibitem#1\endcsname}%
\let\auto@bib@innerbib\@empty
\bibitem [{\citenamefont {Rapoport}(1967)}]{Rapoport1967}%
  \BibitemOpen
  \bibfield  {author} {\bibinfo {author} {\bibfnamefont {E.}~\bibnamefont
  {Rapoport}},\ }\href {\doibase 10.1063/1.1841150} {\bibfield  {journal}
  {\bibinfo  {journal} {J. Chem. Phys.}\ }\textbf {\bibinfo {volume} {46}},\
  \bibinfo {pages} {2891} (\bibinfo {year} {1967})}\BibitemShut {NoStop}%
\bibitem [{\citenamefont {McMillan}\ \emph {et~al.}(2007)\citenamefont
  {McMillan}, \citenamefont {Wilson}, \citenamefont {Wilding}, \citenamefont
  {Daisenberger}, \citenamefont {Mezouar},\ and\ \citenamefont
  {Greaves}}]{McMillan2007}%
  \BibitemOpen
  \bibfield  {author} {\bibinfo {author} {\bibfnamefont {P.~F.}\ \bibnamefont
  {McMillan}}, \bibinfo {author} {\bibfnamefont {M.}~\bibnamefont {Wilson}},
  \bibinfo {author} {\bibfnamefont {M.~C.}\ \bibnamefont {Wilding}}, \bibinfo
  {author} {\bibfnamefont {D.}~\bibnamefont {Daisenberger}}, \bibinfo {author}
  {\bibfnamefont {M.}~\bibnamefont {Mezouar}}, \ and\ \bibinfo {author}
  {\bibfnamefont {G.~N.}\ \bibnamefont {Greaves}},\ }\href {\doibase
  10.1088/0953-8984/19/41/415101} {\bibfield  {journal} {\bibinfo  {journal}
  {Journal of Physics-condensed Matter}\ }\textbf {\bibinfo {volume} {19}},\
  \bibinfo {pages} {415101} (\bibinfo {year} {2007})}\BibitemShut {NoStop}%
\bibitem [{\citenamefont {Barnes}\ \emph {et~al.}(2009)\citenamefont {Barnes},
  \citenamefont {Skinner}, \citenamefont {Salmon}, \citenamefont {Bytchkov},
  \citenamefont {Pozdnyakova}, \citenamefont {Farmer},\ and\ \citenamefont
  {Fischer}}]{Barnes2009}%
  \BibitemOpen
  \bibfield  {author} {\bibinfo {author} {\bibfnamefont {A.~C.}\ \bibnamefont
  {Barnes}}, \bibinfo {author} {\bibfnamefont {L.~B.}\ \bibnamefont {Skinner}},
  \bibinfo {author} {\bibfnamefont {P.~S.}\ \bibnamefont {Salmon}}, \bibinfo
  {author} {\bibfnamefont {A.}~\bibnamefont {Bytchkov}}, \bibinfo {author}
  {\bibfnamefont {I.}~\bibnamefont {Pozdnyakova}}, \bibinfo {author}
  {\bibfnamefont {T.~O.}\ \bibnamefont {Farmer}}, \ and\ \bibinfo {author}
  {\bibfnamefont {H.~E.}\ \bibnamefont {Fischer}},\ }\href
  {http://link.aps.org/doi/10.1103/PhysRevLett.103.225702} {\bibfield
  {journal} {\bibinfo  {journal} {Phys. Rev. Lett.}\ }\textbf {\bibinfo
  {volume} {103}},\ \bibinfo {pages} {225702} (\bibinfo {year}
  {2009})}\BibitemShut {NoStop}%
\bibitem [{\citenamefont {Greaves}\ \emph {et~al.}(2011)\citenamefont
  {Greaves}, \citenamefont {Wilding}, \citenamefont {Langstaff}, \citenamefont
  {Kargl}, \citenamefont {Hennet}, \citenamefont {Benmore}, \citenamefont
  {Weber}, \citenamefont {Van}, \citenamefont {Maj\'erus},\ and\ \citenamefont
  {McMillan}}]{Greaves2011}%
  \BibitemOpen
  \bibfield  {author} {\bibinfo {author} {\bibfnamefont {G.}~\bibnamefont
  {Greaves}}, \bibinfo {author} {\bibfnamefont {M.}~\bibnamefont {Wilding}},
  \bibinfo {author} {\bibfnamefont {D.}~\bibnamefont {Langstaff}}, \bibinfo
  {author} {\bibfnamefont {F.}~\bibnamefont {Kargl}}, \bibinfo {author}
  {\bibfnamefont {L.}~\bibnamefont {Hennet}}, \bibinfo {author} {\bibfnamefont
  {C.}~\bibnamefont {Benmore}}, \bibinfo {author} {\bibfnamefont
  {J.}~\bibnamefont {Weber}}, \bibinfo {author} {\bibfnamefont {Q.~V.}\
  \bibnamefont {Van}}, \bibinfo {author} {\bibfnamefont {O.}~\bibnamefont
  {Maj\'erus}}, \ and\ \bibinfo {author} {\bibfnamefont {P.}~\bibnamefont
  {McMillan}},\ }\bibfield  {booktitle} {\emph {\bibinfo {booktitle} {6th
  International Discussion Meeting on Relaxation in Complex Systems}},\ }\href
  {http://www.sciencedirect.com/science/article/pii/S0022309310005272}
  {\bibfield  {journal} {\bibinfo  {journal} {J. Non-Cryst. Solids}\ }\textbf
  {\bibinfo {volume} {357}},\ \bibinfo {pages} {435} (\bibinfo {year}
  {2011})}\BibitemShut {NoStop}%
\bibitem [{\citenamefont {Wilding}\ \emph {et~al.}(2013)\citenamefont
  {Wilding}, \citenamefont {Wilson}, \citenamefont {Benmore}, \citenamefont
  {Weber},\ and\ \citenamefont {McMillan}}]{Wilding2013}%
  \BibitemOpen
  \bibfield  {author} {\bibinfo {author} {\bibfnamefont {M.~C.}\ \bibnamefont
  {Wilding}}, \bibinfo {author} {\bibfnamefont {M.}~\bibnamefont {Wilson}},
  \bibinfo {author} {\bibfnamefont {C.~J.}\ \bibnamefont {Benmore}}, \bibinfo
  {author} {\bibfnamefont {J.~K.~R.}\ \bibnamefont {Weber}}, \ and\ \bibinfo
  {author} {\bibfnamefont {P.~F.}\ \bibnamefont {McMillan}},\ }\href
  {http://dx.doi.org/10.1039/C3CP51209F} {\bibfield  {journal} {\bibinfo
  {journal} {Phys. Chem. Chem. Phys.}\ }\textbf {\bibinfo {volume} {15}},\
  \bibinfo {pages} {8589} (\bibinfo {year} {2013})}\BibitemShut {NoStop}%
\bibitem [{\citenamefont {Limmer}\ and\ \citenamefont
  {Chandler}(2011)}]{Limmer2011}%
  \BibitemOpen
  \bibfield  {author} {\bibinfo {author} {\bibfnamefont {D.~T.}\ \bibnamefont
  {Limmer}}\ and\ \bibinfo {author} {\bibfnamefont {D.}~\bibnamefont
  {Chandler}},\ }\href {\doibase 10.1063/1.3643333} {\bibfield  {journal}
  {\bibinfo  {journal} {J. Chem. Phys.}\ }\textbf {\bibinfo {volume} {135}},\
  \bibinfo {eid} {134503} (\bibinfo {year} {2011})}\BibitemShut {NoStop}%
\bibitem [{\citenamefont {Liu}\ \emph {et~al.}(2012)\citenamefont {Liu},
  \citenamefont {Palmer}, \citenamefont {Panagiotopoulos},\ and\ \citenamefont
  {Debenedetti}}]{Liu2012}%
  \BibitemOpen
  \bibfield  {author} {\bibinfo {author} {\bibfnamefont {Y.}~\bibnamefont
  {Liu}}, \bibinfo {author} {\bibfnamefont {J.~C.}\ \bibnamefont {Palmer}},
  \bibinfo {author} {\bibfnamefont {A.~Z.}\ \bibnamefont {Panagiotopoulos}}, \
  and\ \bibinfo {author} {\bibfnamefont {P.~G.}\ \bibnamefont {Debenedetti}},\
  }\href {\doibase 10.1063/1.4769126} {\bibfield  {journal} {\bibinfo
  {journal} {J. Chem. Phys.}\ }\textbf {\bibinfo {volume} {137}},\ \bibinfo
  {eid} {214505} (\bibinfo {year} {2012})}\BibitemShut {NoStop}%
\bibitem [{\citenamefont {Katayama}\ \emph {et~al.}(2000)\citenamefont
  {Katayama}, \citenamefont {Mizutani}, \citenamefont {Utsumi}, \citenamefont
  {Shimomura}, \citenamefont {Yamakata},\ and\ \citenamefont
  {Funakoshi}}]{Katayama2000}%
  \BibitemOpen
  \bibfield  {author} {\bibinfo {author} {\bibfnamefont {Y.}~\bibnamefont
  {Katayama}}, \bibinfo {author} {\bibfnamefont {T.}~\bibnamefont {Mizutani}},
  \bibinfo {author} {\bibfnamefont {W.}~\bibnamefont {Utsumi}}, \bibinfo
  {author} {\bibfnamefont {O.}~\bibnamefont {Shimomura}}, \bibinfo {author}
  {\bibfnamefont {M.}~\bibnamefont {Yamakata}}, \ and\ \bibinfo {author}
  {\bibfnamefont {K.-i.}\ \bibnamefont {Funakoshi}},\ }\href
  {http://dx.doi.org/10.1038/35003143} {\bibfield  {journal} {\bibinfo
  {journal} {Nature}\ }\textbf {\bibinfo {volume} {403}},\ \bibinfo {pages}
  {170} (\bibinfo {year} {2000})}\BibitemShut {NoStop}%
\bibitem [{\citenamefont {Aasland}\ and\ \citenamefont
  {McMillan}(1994)}]{Aasland1994}%
  \BibitemOpen
  \bibfield  {author} {\bibinfo {author} {\bibfnamefont {S.}~\bibnamefont
  {Aasland}}\ and\ \bibinfo {author} {\bibfnamefont {P.}~\bibnamefont
  {McMillan}},\ }\href {http://dx.doi.org/10.1038/369633a0} {\bibfield
  {journal} {\bibinfo  {journal} {Nature}\ }\textbf {\bibinfo {volume} {369}},\
  \bibinfo {pages} {633} (\bibinfo {year} {1994})}\BibitemShut {NoStop}%
\bibitem [{\citenamefont {Greaves}\ \emph {et~al.}(2008)\citenamefont
  {Greaves}, \citenamefont {Wilding}, \citenamefont {Fearn}, \citenamefont
  {Langstaff}, \citenamefont {Kargl}, \citenamefont {Cox}, \citenamefont {Van},
  \citenamefont {Majerus}, \citenamefont {Benmore}, \citenamefont {Weber},
  \citenamefont {Martin},\ and\ \citenamefont {Hennet}}]{Greaves2008}%
  \BibitemOpen
  \bibfield  {author} {\bibinfo {author} {\bibfnamefont {G.~N.}\ \bibnamefont
  {Greaves}}, \bibinfo {author} {\bibfnamefont {M.~C.}\ \bibnamefont
  {Wilding}}, \bibinfo {author} {\bibfnamefont {S.}~\bibnamefont {Fearn}},
  \bibinfo {author} {\bibfnamefont {D.}~\bibnamefont {Langstaff}}, \bibinfo
  {author} {\bibfnamefont {F.}~\bibnamefont {Kargl}}, \bibinfo {author}
  {\bibfnamefont {S.}~\bibnamefont {Cox}}, \bibinfo {author} {\bibfnamefont
  {Q.~V.}\ \bibnamefont {Van}}, \bibinfo {author} {\bibfnamefont
  {O.}~\bibnamefont {Majerus}}, \bibinfo {author} {\bibfnamefont {C.~J.}\
  \bibnamefont {Benmore}}, \bibinfo {author} {\bibfnamefont {R.}~\bibnamefont
  {Weber}}, \bibinfo {author} {\bibfnamefont {C.~M.}\ \bibnamefont {Martin}}, \
  and\ \bibinfo {author} {\bibfnamefont {L.}~\bibnamefont {Hennet}},\ }\href
  {\doibase 10.1126/science.1160766} {\bibfield  {journal} {\bibinfo  {journal}
  {Science}\ }\textbf {\bibinfo {volume} {322}},\ \bibinfo {pages} {566}
  (\bibinfo {year} {2008})}\BibitemShut {NoStop}%
\bibitem [{\citenamefont {Tanaka}\ \emph {et~al.}(2004)\citenamefont {Tanaka},
  \citenamefont {Kurita},\ and\ \citenamefont {Mataki}}]{Tanaka2004}%
  \BibitemOpen
  \bibfield  {author} {\bibinfo {author} {\bibfnamefont {H.}~\bibnamefont
  {Tanaka}}, \bibinfo {author} {\bibfnamefont {R.}~\bibnamefont {Kurita}}, \
  and\ \bibinfo {author} {\bibfnamefont {H.}~\bibnamefont {Mataki}},\ }\href
  {http://link.aps.org/doi/10.1103/PhysRevLett.92.025701} {\bibfield  {journal}
  {\bibinfo  {journal} {Phys. Rev. Lett.}\ }\textbf {\bibinfo {volume} {92}},\
  \bibinfo {pages} {025701} (\bibinfo {year} {2004})}\BibitemShut {NoStop}%
\bibitem [{\citenamefont {Kurita}\ and\ \citenamefont
  {Tanaka}(2004)}]{Kurita2004}%
  \BibitemOpen
  \bibfield  {author} {\bibinfo {author} {\bibfnamefont {R.}~\bibnamefont
  {Kurita}}\ and\ \bibinfo {author} {\bibfnamefont {H.}~\bibnamefont
  {Tanaka}},\ }\href {\doibase 10.1126/science.1103073} {\bibfield  {journal}
  {\bibinfo  {journal} {Science}\ }\textbf {\bibinfo {volume} {306}},\ \bibinfo
  {pages} {845} (\bibinfo {year} {2004})}\BibitemShut {NoStop}%
\bibitem [{\citenamefont {Kurita}\ and\ \citenamefont
  {Tanaka}(2005)}]{Kurita2005}%
  \BibitemOpen
  \bibfield  {author} {\bibinfo {author} {\bibfnamefont {R.}~\bibnamefont
  {Kurita}}\ and\ \bibinfo {author} {\bibfnamefont {H.}~\bibnamefont
  {Tanaka}},\ }\href {http://stacks.iop.org/0953-8984/17/i=27/a=L01} {\bibfield
   {journal} {\bibinfo  {journal} {J. Phys.: Condens. Matter}\ }\textbf
  {\bibinfo {volume} {17}},\ \bibinfo {pages} {L293} (\bibinfo {year}
  {2005})}\BibitemShut {NoStop}%
\bibitem [{\citenamefont {Brazhkin}\ \emph {et~al.}(2008)\citenamefont
  {Brazhkin}, \citenamefont {Katayama}, \citenamefont {Kondrin}, \citenamefont
  {Hattori}, \citenamefont {Lyapin},\ and\ \citenamefont
  {Saitoh}}]{Brazhkin2008}%
  \BibitemOpen
  \bibfield  {author} {\bibinfo {author} {\bibfnamefont {V.~V.}\ \bibnamefont
  {Brazhkin}}, \bibinfo {author} {\bibfnamefont {Y.}~\bibnamefont {Katayama}},
  \bibinfo {author} {\bibfnamefont {M.~V.}\ \bibnamefont {Kondrin}}, \bibinfo
  {author} {\bibfnamefont {T.}~\bibnamefont {Hattori}}, \bibinfo {author}
  {\bibfnamefont {A.~G.}\ \bibnamefont {Lyapin}}, \ and\ \bibinfo {author}
  {\bibfnamefont {H.}~\bibnamefont {Saitoh}},\ }\href
  {http://link.aps.org/doi/10.1103/PhysRevLett.100.145701} {\bibfield
  {journal} {\bibinfo  {journal} {Phys. Rev. Lett.}\ }\textbf {\bibinfo
  {volume} {100}},\ \bibinfo {pages} {145701} (\bibinfo {year}
  {2008})}\BibitemShut {NoStop}%
\bibitem [{\citenamefont {Brazhkin}\ \emph {et~al.}(2009)\citenamefont
  {Brazhkin}, \citenamefont {Kanzaki}, \citenamefont {Funakoshi},\ and\
  \citenamefont {Katayama}}]{Brazhkin2009}%
  \BibitemOpen
  \bibfield  {author} {\bibinfo {author} {\bibfnamefont {V.~V.}\ \bibnamefont
  {Brazhkin}}, \bibinfo {author} {\bibfnamefont {M.}~\bibnamefont {Kanzaki}},
  \bibinfo {author} {\bibfnamefont {K.-i.}\ \bibnamefont {Funakoshi}}, \ and\
  \bibinfo {author} {\bibfnamefont {Y.}~\bibnamefont {Katayama}},\ }\href
  {http://link.aps.org/doi/10.1103/PhysRevLett.102.115901} {\bibfield
  {journal} {\bibinfo  {journal} {Phys. Rev. Lett.}\ }\textbf {\bibinfo
  {volume} {102}},\ \bibinfo {pages} {115901} (\bibinfo {year}
  {2009})}\BibitemShut {NoStop}%
\bibitem [{\citenamefont {Poole}\ \emph {et~al.}(1992)\citenamefont {Poole},
  \citenamefont {Sciortino}, \citenamefont {Essmann},\ and\ \citenamefont
  {Stanley}}]{Poole1992}%
  \BibitemOpen
  \bibfield  {author} {\bibinfo {author} {\bibfnamefont {P.~H.}\ \bibnamefont
  {Poole}}, \bibinfo {author} {\bibfnamefont {F.}~\bibnamefont {Sciortino}},
  \bibinfo {author} {\bibfnamefont {U.}~\bibnamefont {Essmann}}, \ and\
  \bibinfo {author} {\bibfnamefont {H.~E.}\ \bibnamefont {Stanley}},\ }\href
  {http://dx.doi.org/10.1038/360324a0} {\bibfield  {journal} {\bibinfo
  {journal} {Nature}\ }\textbf {\bibinfo {volume} {360}},\ \bibinfo {pages}
  {324} (\bibinfo {year} {1992})}\BibitemShut {NoStop}%
\bibitem [{\citenamefont {Sciortino}\ \emph {et~al.}(1997)\citenamefont
  {Sciortino}, \citenamefont {Poole}, \citenamefont {Essmann},\ and\
  \citenamefont {Stanley}}]{Sciortino1997}%
  \BibitemOpen
  \bibfield  {author} {\bibinfo {author} {\bibfnamefont {F.}~\bibnamefont
  {Sciortino}}, \bibinfo {author} {\bibfnamefont {P.~H.}\ \bibnamefont
  {Poole}}, \bibinfo {author} {\bibfnamefont {U.}~\bibnamefont {Essmann}}, \
  and\ \bibinfo {author} {\bibfnamefont {H.~E.}\ \bibnamefont {Stanley}},\
  }\href {http://link.aps.org/doi/10.1103/PhysRevE.55.727} {\bibfield
  {journal} {\bibinfo  {journal} {Phys. Rev. E}\ }\textbf {\bibinfo {volume}
  {55}},\ \bibinfo {pages} {727} (\bibinfo {year} {1997})}\BibitemShut
  {NoStop}%
\bibitem [{\citenamefont {Liu}\ \emph {et~al.}(2005)\citenamefont {Liu},
  \citenamefont {Chen}, \citenamefont {Faraone}, \citenamefont {Yen},\ and\
  \citenamefont {Mou}}]{Liu2005}%
  \BibitemOpen
  \bibfield  {author} {\bibinfo {author} {\bibfnamefont {L.}~\bibnamefont
  {Liu}}, \bibinfo {author} {\bibfnamefont {S.-H.}\ \bibnamefont {Chen}},
  \bibinfo {author} {\bibfnamefont {A.}~\bibnamefont {Faraone}}, \bibinfo
  {author} {\bibfnamefont {C.-W.}\ \bibnamefont {Yen}}, \ and\ \bibinfo
  {author} {\bibfnamefont {C.-Y.}\ \bibnamefont {Mou}},\ }\href
  {http://link.aps.org/doi/10.1103/PhysRevLett.95.117802} {\bibfield  {journal}
  {\bibinfo  {journal} {Phys. Rev. Lett.}\ }\textbf {\bibinfo {volume} {95}},\
  \bibinfo {pages} {117802} (\bibinfo {year} {2005})}\BibitemShut {NoStop}%
\bibitem [{\citenamefont {Mallamace}\ \emph {et~al.}(2008)\citenamefont
  {Mallamace}, \citenamefont {Branca}, \citenamefont {Broccio}, \citenamefont
  {Corsaro}, \citenamefont {Gonzalez-Segredo}, \citenamefont {Spooren},
  \citenamefont {Stanley},\ and\ \citenamefont {Chen}}]{Mallamace2008}%
  \BibitemOpen
  \bibfield  {author} {\bibinfo {author} {\bibfnamefont {F.}~\bibnamefont
  {Mallamace}}, \bibinfo {author} {\bibfnamefont {C.}~\bibnamefont {Branca}},
  \bibinfo {author} {\bibfnamefont {M.}~\bibnamefont {Broccio}}, \bibinfo
  {author} {\bibfnamefont {C.}~\bibnamefont {Corsaro}}, \bibinfo {author}
  {\bibfnamefont {N.}~\bibnamefont {Gonzalez-Segredo}}, \bibinfo {author}
  {\bibfnamefont {J.}~\bibnamefont {Spooren}}, \bibinfo {author} {\bibfnamefont
  {H.~E.}\ \bibnamefont {Stanley}}, \ and\ \bibinfo {author} {\bibfnamefont
  {S.~H.}\ \bibnamefont {Chen}},\ }\href {\doibase 10.1140/epjst/e2008-00747-2}
  {\bibfield  {journal} {\bibinfo  {journal} {European Physical Journal-special
  Topics}\ }\textbf {\bibinfo {volume} {161}},\ \bibinfo {pages} {19} (\bibinfo
  {year} {2008})}\BibitemShut {NoStop}%
\bibitem [{\citenamefont {Zhang}\ \emph {et~al.}(2011)\citenamefont {Zhang},
  \citenamefont {Faraone}, \citenamefont {Kamitakahara}, \citenamefont {Liu},
  \citenamefont {Mou}, \citenamefont {Le\~{a}o}, \citenamefont {Chang},\ and\
  \citenamefont {Chen}}]{Zhang2011}%
  \BibitemOpen
  \bibfield  {author} {\bibinfo {author} {\bibfnamefont {Y.}~\bibnamefont
  {Zhang}}, \bibinfo {author} {\bibfnamefont {A.}~\bibnamefont {Faraone}},
  \bibinfo {author} {\bibfnamefont {W.~A.}\ \bibnamefont {Kamitakahara}},
  \bibinfo {author} {\bibfnamefont {K.-H.}\ \bibnamefont {Liu}}, \bibinfo
  {author} {\bibfnamefont {C.-Y.}\ \bibnamefont {Mou}}, \bibinfo {author}
  {\bibfnamefont {J.~B.}\ \bibnamefont {Le\~{a}o}}, \bibinfo {author}
  {\bibfnamefont {S.}~\bibnamefont {Chang}}, \ and\ \bibinfo {author}
  {\bibfnamefont {S.-H.}\ \bibnamefont {Chen}},\ }\href {\doibase
  10.1073/pnas.1100238108} {\bibfield  {journal} {\bibinfo  {journal} {Proc.
  Natl. Acad. Sci. U.S.A.}\ }\textbf {\bibinfo {volume} {108}},\ \bibinfo
  {pages} {12206} (\bibinfo {year} {2011})}\BibitemShut {NoStop}%
\bibitem [{\citenamefont {Sastry}\ and\ \citenamefont
  {Angell}(2003)}]{Sastry2003}%
  \BibitemOpen
  \bibfield  {author} {\bibinfo {author} {\bibfnamefont {S.}~\bibnamefont
  {Sastry}}\ and\ \bibinfo {author} {\bibfnamefont {C.~A.}\ \bibnamefont
  {Angell}},\ }\href {\doibase 10.1038/nmat994} {\bibfield  {journal} {\bibinfo
   {journal} {Nat. Mater.}\ }\textbf {\bibinfo {volume} {2}},\ \bibinfo {pages}
  {739} (\bibinfo {year} {2003})}\BibitemShut {NoStop}%
\bibitem [{\citenamefont {Miranda}\ and\ \citenamefont
  {Antonelli}(2004)}]{Miranda2004}%
  \BibitemOpen
  \bibfield  {author} {\bibinfo {author} {\bibfnamefont {C.~R.}\ \bibnamefont
  {Miranda}}\ and\ \bibinfo {author} {\bibfnamefont {A.}~\bibnamefont
  {Antonelli}},\ }\href {\doibase 10.1063/1.1755653} {\bibfield  {journal}
  {\bibinfo  {journal} {J. Chem. Phys.}\ }\textbf {\bibinfo {volume} {120}},\
  \bibinfo {pages} {11672} (\bibinfo {year} {2004})}\BibitemShut {NoStop}%
\bibitem [{\citenamefont {Jakse}\ and\ \citenamefont
  {Pasturel}(2007)}]{Jakse2007}%
  \BibitemOpen
  \bibfield  {author} {\bibinfo {author} {\bibfnamefont {N.}~\bibnamefont
  {Jakse}}\ and\ \bibinfo {author} {\bibfnamefont {A.}~\bibnamefont
  {Pasturel}},\ }\href {\doibase 10.1103/PhysRevLett.99.205702} {\bibfield
  {journal} {\bibinfo  {journal} {Phys. Rev. Lett.}\ }\textbf {\bibinfo
  {volume} {99}},\ \bibinfo {pages} {205702} (\bibinfo {year}
  {2007})}\BibitemShut {NoStop}%
\bibitem [{\citenamefont {Ganesh}\ and\ \citenamefont
  {Widom}(2009)}]{Ganesh2009}%
  \BibitemOpen
  \bibfield  {author} {\bibinfo {author} {\bibfnamefont {P.}~\bibnamefont
  {Ganesh}}\ and\ \bibinfo {author} {\bibfnamefont {M.}~\bibnamefont {Widom}},\
  }\href {\doibase 10.1103/PhysRevLett.102.075701} {\bibfield  {journal}
  {\bibinfo  {journal} {Phys. Rev. Lett.}\ }\textbf {\bibinfo {volume} {102}},\
  \bibinfo {pages} {075701} (\bibinfo {year} {2009})}\BibitemShut {NoStop}%
\bibitem [{\citenamefont {Garcez}\ and\ \citenamefont
  {Antonelli}(2011)}]{Garcez2011}%
  \BibitemOpen
  \bibfield  {author} {\bibinfo {author} {\bibfnamefont {K.~M.~S.}\
  \bibnamefont {Garcez}}\ and\ \bibinfo {author} {\bibfnamefont
  {A.}~\bibnamefont {Antonelli}},\ }\href {http://dx.doi.org/10.1063/1.3663387}
  {\bibfield  {journal} {\bibinfo  {journal} {J. Chem. Phys.}\ }\textbf
  {\bibinfo {volume} {135}},\ \bibinfo {pages} {204508} (\bibinfo {year}
  {2011})}\BibitemShut {NoStop}%
\bibitem [{\citenamefont {Tien}\ \emph {et~al.}(2006)\citenamefont {Tien},
  \citenamefont {Charnaya}, \citenamefont {Wang}, \citenamefont {Kumzerov},\
  and\ \citenamefont {Michel}}]{Tien2006}%
  \BibitemOpen
  \bibfield  {author} {\bibinfo {author} {\bibfnamefont {C.}~\bibnamefont
  {Tien}}, \bibinfo {author} {\bibfnamefont {E.~V.}\ \bibnamefont {Charnaya}},
  \bibinfo {author} {\bibfnamefont {W.}~\bibnamefont {Wang}}, \bibinfo {author}
  {\bibfnamefont {Y.~A.}\ \bibnamefont {Kumzerov}}, \ and\ \bibinfo {author}
  {\bibfnamefont {D.}~\bibnamefont {Michel}},\ }\href
  {http://link.aps.org/doi/10.1103/PhysRevB.74.024116} {\bibfield  {journal}
  {\bibinfo  {journal} {Phys. Rev. B}\ }\textbf {\bibinfo {volume} {74}},\
  \bibinfo {pages} {024116} (\bibinfo {year} {2006})}\BibitemShut {NoStop}%
\bibitem [{\citenamefont {Jara}\ \emph {et~al.}(2009)\citenamefont {Jara},
  \citenamefont {Michelon}, \citenamefont {Antonelli},\ and\ \citenamefont
  {de~Koning}}]{Jara2009}%
  \BibitemOpen
  \bibfield  {author} {\bibinfo {author} {\bibfnamefont {D.~A.~C.}\
  \bibnamefont {Jara}}, \bibinfo {author} {\bibfnamefont {M.~F.}\ \bibnamefont
  {Michelon}}, \bibinfo {author} {\bibfnamefont {A.}~\bibnamefont {Antonelli}},
  \ and\ \bibinfo {author} {\bibfnamefont {M.}~\bibnamefont {de~Koning}},\
  }\href {\doibase 10.1063/1.3154424} {\bibfield  {journal} {\bibinfo
  {journal} {J. Chem. Phys.}\ }\textbf {\bibinfo {volume} {130}},\ \bibinfo
  {pages} {221101} (\bibinfo {year} {2009})}\BibitemShut {NoStop}%
\bibitem [{\citenamefont {Boates}\ and\ \citenamefont
  {Bonev}(2009)}]{Boates2009}%
  \BibitemOpen
  \bibfield  {author} {\bibinfo {author} {\bibfnamefont {B.}~\bibnamefont
  {Boates}}\ and\ \bibinfo {author} {\bibfnamefont {S.~A.}\ \bibnamefont
  {Bonev}},\ }\href {http://link.aps.org/doi/10.1103/PhysRevLett.102.015701}
  {\bibfield  {journal} {\bibinfo  {journal} {Phys. Rev. Lett.}\ }\textbf
  {\bibinfo {volume} {102}},\ \bibinfo {pages} {015701} (\bibinfo {year}
  {2009})}\BibitemShut {NoStop}%
\bibitem [{\citenamefont {Poole}\ \emph {et~al.}(2011)\citenamefont {Poole},
  \citenamefont {Becker}, \citenamefont {Sciortino},\ and\ \citenamefont
  {Starr}}]{Poole2011}%
  \BibitemOpen
  \bibfield  {author} {\bibinfo {author} {\bibfnamefont {P.~H.}\ \bibnamefont
  {Poole}}, \bibinfo {author} {\bibfnamefont {S.~R.}\ \bibnamefont {Becker}},
  \bibinfo {author} {\bibfnamefont {F.}~\bibnamefont {Sciortino}}, \ and\
  \bibinfo {author} {\bibfnamefont {F.~W.}\ \bibnamefont {Starr}},\ }\href
  {\doibase 10.1021/jp204889m} {\bibfield  {journal} {\bibinfo  {journal} {J.
  Phys. Chem. B}\ }\textbf {\bibinfo {volume} {115}},\ \bibinfo {pages} {14176}
  (\bibinfo {year} {2011})}\BibitemShut {NoStop}%
\bibitem [{\citenamefont {Cajahuaringa}\ \emph {et~al.}(2012)\citenamefont
  {Cajahuaringa}, \citenamefont {de~Koning},\ and\ \citenamefont
  {Antonelli}}]{Cajahuaringa2012}%
  \BibitemOpen
  \bibfield  {author} {\bibinfo {author} {\bibfnamefont {S.}~\bibnamefont
  {Cajahuaringa}}, \bibinfo {author} {\bibfnamefont {M.}~\bibnamefont
  {de~Koning}}, \ and\ \bibinfo {author} {\bibfnamefont {A.}~\bibnamefont
  {Antonelli}},\ }\href {http://dx.doi.org/10.1063/1.3684550} {\bibfield
  {journal} {\bibinfo  {journal} {J. Chem. Phys.}\ }\textbf {\bibinfo {volume}
  {136}},\ \bibinfo {pages} {064513} (\bibinfo {year} {2012})}\BibitemShut
  {NoStop}%
\bibitem [{\citenamefont {Lad}\ \emph {et~al.}(2012)\citenamefont {Lad},
  \citenamefont {Jakse},\ and\ \citenamefont {Pasturel}}]{Lad2012}%
  \BibitemOpen
  \bibfield  {author} {\bibinfo {author} {\bibfnamefont {K.~N.}\ \bibnamefont
  {Lad}}, \bibinfo {author} {\bibfnamefont {N.}~\bibnamefont {Jakse}}, \ and\
  \bibinfo {author} {\bibfnamefont {A.}~\bibnamefont {Pasturel}},\ }\href
  {http://dx.doi.org/10.1063/1.3692610} {\bibfield  {journal} {\bibinfo
  {journal} {J. Chem. Phys.}\ }\textbf {\bibinfo {volume} {136}},\ \bibinfo
  {pages} {104509} (\bibinfo {year} {2012})}\BibitemShut {NoStop}%
\bibitem [{\citenamefont {Malinovsky}\ \emph {et~al.}(1990)\citenamefont
  {Malinovsky}, \citenamefont {Novikov}, \citenamefont {Parshin}, \citenamefont
  {Sokolov},\ and\ \citenamefont {Zemlyanov}}]{Malinovsky1990}%
  \BibitemOpen
  \bibfield  {author} {\bibinfo {author} {\bibfnamefont {V.~K.}\ \bibnamefont
  {Malinovsky}}, \bibinfo {author} {\bibfnamefont {V.~N.}\ \bibnamefont
  {Novikov}}, \bibinfo {author} {\bibfnamefont {P.~P.}\ \bibnamefont
  {Parshin}}, \bibinfo {author} {\bibfnamefont {A.~P.}\ \bibnamefont
  {Sokolov}}, \ and\ \bibinfo {author} {\bibfnamefont {M.~G.}\ \bibnamefont
  {Zemlyanov}},\ }\href {http://stacks.iop.org/0295-5075/11/i=1/a=008}
  {\bibfield  {journal} {\bibinfo  {journal} {Europhys. Lett.}\ }\textbf
  {\bibinfo {volume} {11}},\ \bibinfo {pages} {43} (\bibinfo {year}
  {1990})}\BibitemShut {NoStop}%
\bibitem [{\citenamefont {Nakayama}(2002)}]{Nakayama2002}%
  \BibitemOpen
  \bibfield  {author} {\bibinfo {author} {\bibfnamefont {T.}~\bibnamefont
  {Nakayama}},\ }\href {http://stacks.iop.org/0034-4885/65/i=8/a=203}
  {\bibfield  {journal} {\bibinfo  {journal} {Reports on Progress in Physics}\
  }\textbf {\bibinfo {volume} {65}},\ \bibinfo {pages} {1195} (\bibinfo {year}
  {2002})}\BibitemShut {NoStop}%
\bibitem [{\citenamefont {Binder}\ and\ \citenamefont
  {Kob}(2011)}]{Binder2011}%
  \BibitemOpen
  \bibfield  {author} {\bibinfo {author} {\bibfnamefont {K.}~\bibnamefont
  {Binder}}\ and\ \bibinfo {author} {\bibfnamefont {W.}~\bibnamefont {Kob}},\
  }\href {http://books.google.com.br/books?id=lEOWIZeFSP4C} {\emph {\bibinfo
  {title} {Glassy Materials and Disordered Solids: An Introduction to Their
  Statistical Mechanics}}}\ (\bibinfo  {publisher} {World Scientific},\
  \bibinfo {year} {2011})\BibitemShut {NoStop}%
\bibitem [{\citenamefont {Jakse}\ \emph {et~al.}(2009)\citenamefont {Jakse},
  \citenamefont {Pasturel}, \citenamefont {Sastry},\ and\ \citenamefont
  {Angell}}]{Jakse2009}%
  \BibitemOpen
  \bibfield  {author} {\bibinfo {author} {\bibfnamefont {N.}~\bibnamefont
  {Jakse}}, \bibinfo {author} {\bibfnamefont {A.}~\bibnamefont {Pasturel}},
  \bibinfo {author} {\bibfnamefont {S.}~\bibnamefont {Sastry}}, \ and\ \bibinfo
  {author} {\bibfnamefont {C.~A.}\ \bibnamefont {Angell}},\ }\href
  {http://dx.doi.org/10.1063/1.3154368} {\bibfield  {journal} {\bibinfo
  {journal} {J. Chem. Phys.}\ }\textbf {\bibinfo {volume} {130}},\ \bibinfo
  {pages} {247103} (\bibinfo {year} {2009})}\BibitemShut {NoStop}%
\bibitem [{\citenamefont {Yannopoulos}(2009)}]{Yannopoulos2009}%
  \BibitemOpen
  \bibfield  {author} {\bibinfo {author} {\bibfnamefont {S.~N.}\ \bibnamefont
  {Yannopoulos}},\ }\href {http://dx.doi.org/10.1063/1.3151970} {\bibfield
  {journal} {\bibinfo  {journal} {J. Chem. Phys.}\ }\textbf {\bibinfo {volume}
  {130}},\ \bibinfo {pages} {247102} (\bibinfo {year} {2009})}\BibitemShut
  {NoStop}%
\bibitem [{\citenamefont {Sokolov}\ \emph {et~al.}(1993)\citenamefont
  {Sokolov}, \citenamefont {R\"{o}ssler}, \citenamefont {Kisliuk},\ and\
  \citenamefont {Quitmann}}]{Sokolov1993}%
  \BibitemOpen
  \bibfield  {author} {\bibinfo {author} {\bibfnamefont {A.~P.}\ \bibnamefont
  {Sokolov}}, \bibinfo {author} {\bibfnamefont {E.}~\bibnamefont
  {R\"{o}ssler}}, \bibinfo {author} {\bibfnamefont {A.}~\bibnamefont
  {Kisliuk}}, \ and\ \bibinfo {author} {\bibfnamefont {D.}~\bibnamefont
  {Quitmann}},\ }\href {http://link.aps.org/doi/10.1103/PhysRevLett.71.2062}
  {\bibfield  {journal} {\bibinfo  {journal} {Phys. Rev. Lett.}\ }\textbf
  {\bibinfo {volume} {71}},\ \bibinfo {pages} {2062} (\bibinfo {year}
  {1993})}\BibitemShut {NoStop}%
\bibitem [{\citenamefont {R\"{o}ssler}\ and\ \citenamefont
  {Sokolov}(1996)}]{Rossler1996}%
  \BibitemOpen
  \bibfield  {author} {\bibinfo {author} {\bibfnamefont {E.}~\bibnamefont
  {R\"{o}ssler}}\ and\ \bibinfo {author} {\bibfnamefont {A.}~\bibnamefont
  {Sokolov}},\ }\href {\doibase 10.1016/0009-2541(95)00169-7} {\bibfield
  {journal} {\bibinfo  {journal} {Chem. Geol.}\ }\textbf {\bibinfo {volume}
  {128}},\ \bibinfo {pages} {143 } (\bibinfo {year} {1996})},\ \bibinfo {note}
  {<ce:title>5TH Silicate Melt Workshop</ce:title>}\BibitemShut {NoStop}%
\bibitem [{\citenamefont {Sokolov}\ \emph {et~al.}(1997)\citenamefont
  {Sokolov}, \citenamefont {Calemczuk}, \citenamefont {Salce}, \citenamefont
  {Kisliuk}, \citenamefont {Quitmann},\ and\ \citenamefont
  {Duval}}]{Sokolov1997}%
  \BibitemOpen
  \bibfield  {author} {\bibinfo {author} {\bibfnamefont {A.~P.}\ \bibnamefont
  {Sokolov}}, \bibinfo {author} {\bibfnamefont {R.}~\bibnamefont {Calemczuk}},
  \bibinfo {author} {\bibfnamefont {B.}~\bibnamefont {Salce}}, \bibinfo
  {author} {\bibfnamefont {A.}~\bibnamefont {Kisliuk}}, \bibinfo {author}
  {\bibfnamefont {D.}~\bibnamefont {Quitmann}}, \ and\ \bibinfo {author}
  {\bibfnamefont {E.}~\bibnamefont {Duval}},\ }\href
  {http://link.aps.org/doi/10.1103/PhysRevLett.78.2405} {\bibfield  {journal}
  {\bibinfo  {journal} {Phys. Rev. Lett.}\ }\textbf {\bibinfo {volume} {78}},\
  \bibinfo {pages} {2405} (\bibinfo {year} {1997})}\BibitemShut {NoStop}%
\bibitem [{\citenamefont {Novikov}\ \emph {et~al.}(2005)\citenamefont
  {Novikov}, \citenamefont {Ding},\ and\ \citenamefont
  {Sokolov}}]{Novikov2005}%
  \BibitemOpen
  \bibfield  {author} {\bibinfo {author} {\bibfnamefont {V.~N.}\ \bibnamefont
  {Novikov}}, \bibinfo {author} {\bibfnamefont {Y.}~\bibnamefont {Ding}}, \
  and\ \bibinfo {author} {\bibfnamefont {A.~P.}\ \bibnamefont {Sokolov}},\
  }\href {http://link.aps.org/doi/10.1103/PhysRevE.71.061501} {\bibfield
  {journal} {\bibinfo  {journal} {Phys. Rev. E}\ }\textbf {\bibinfo {volume}
  {71}},\ \bibinfo {pages} {061501} (\bibinfo {year} {2005})}\BibitemShut
  {NoStop}%
\bibitem [{\citenamefont {Horbach}\ \emph {et~al.}(1996)\citenamefont
  {Horbach}, \citenamefont {Kob}, \citenamefont {Binder},\ and\ \citenamefont
  {Angell}}]{Horbach1996}%
  \BibitemOpen
  \bibfield  {author} {\bibinfo {author} {\bibfnamefont {J.}~\bibnamefont
  {Horbach}}, \bibinfo {author} {\bibfnamefont {W.}~\bibnamefont {Kob}},
  \bibinfo {author} {\bibfnamefont {K.}~\bibnamefont {Binder}}, \ and\ \bibinfo
  {author} {\bibfnamefont {C.~A.}\ \bibnamefont {Angell}},\ }\href
  {http://link.aps.org/doi/10.1103/PhysRevE.54.R5897} {\bibfield  {journal}
  {\bibinfo  {journal} {Phys. Rev. E}\ }\textbf {\bibinfo {volume} {54}},\
  \bibinfo {pages} {R5897} (\bibinfo {year} {1996})}\BibitemShut {NoStop}%
\bibitem [{\citenamefont {Stillinger}\ and\ \citenamefont
  {Weber}(1985)}]{Stillinger1985}%
  \BibitemOpen
  \bibfield  {author} {\bibinfo {author} {\bibfnamefont {F.~H.}\ \bibnamefont
  {Stillinger}}\ and\ \bibinfo {author} {\bibfnamefont {T.~A.}\ \bibnamefont
  {Weber}},\ }\href {http://link.aps.org/doi/10.1103/PhysRevB.31.5262}
  {\bibfield  {journal} {\bibinfo  {journal} {Phys. Rev. B}\ }\textbf {\bibinfo
  {volume} {31}},\ \bibinfo {pages} {5262} (\bibinfo {year}
  {1985})}\BibitemShut {NoStop}%
\bibitem [{\citenamefont {Jakse}\ and\ \citenamefont
  {Pasturel}(2008)}]{Jakse2008}%
  \BibitemOpen
  \bibfield  {author} {\bibinfo {author} {\bibfnamefont {N.}~\bibnamefont
  {Jakse}}\ and\ \bibinfo {author} {\bibfnamefont {A.}~\bibnamefont
  {Pasturel}},\ }\href {http://dx.doi.org/10.1063/1.2970084} {\bibfield
  {journal} {\bibinfo  {journal} {J. Chem. Phys.}\ }\textbf {\bibinfo {volume}
  {129}},\ \bibinfo {pages} {104503} (\bibinfo {year} {2008})}\BibitemShut
  {NoStop}%
\bibitem [{\citenamefont {Yannopoulos}\ and\ \citenamefont
  {Papatheodorou}(2000)}]{Yannopoulos2000}%
  \BibitemOpen
  \bibfield  {author} {\bibinfo {author} {\bibfnamefont {S.~N.}\ \bibnamefont
  {Yannopoulos}}\ and\ \bibinfo {author} {\bibfnamefont {G.~N.}\ \bibnamefont
  {Papatheodorou}},\ }\href {http://link.aps.org/doi/10.1103/PhysRevB.62.3728}
  {\bibfield  {journal} {\bibinfo  {journal} {Phys. Rev. B}\ }\textbf {\bibinfo
  {volume} {62}},\ \bibinfo {pages} {3728} (\bibinfo {year}
  {2000})}\BibitemShut {NoStop}%
\bibitem [{\citenamefont {Angell}(1988)}]{Angell1988}%
  \BibitemOpen
  \bibfield  {author} {\bibinfo {author} {\bibfnamefont {C.~A.}\ \bibnamefont
  {Angell}},\ }\href {\doibase 10.1016/0022-3093(88)90133-0} {\bibfield
  {journal} {\bibinfo  {journal} {J. Non-Cryst. Solids}\ }\textbf {\bibinfo
  {volume} {102}},\ \bibinfo {pages} {205} (\bibinfo {year}
  {1988})}\BibitemShut {NoStop}%
\bibitem [{\citenamefont {Baskes}\ \emph {et~al.}(2002)\citenamefont {Baskes},
  \citenamefont {Chen},\ and\ \citenamefont {Cherne}}]{Baskes2002}%
  \BibitemOpen
  \bibfield  {author} {\bibinfo {author} {\bibfnamefont {M.~I.}\ \bibnamefont
  {Baskes}}, \bibinfo {author} {\bibfnamefont {S.~P.}\ \bibnamefont {Chen}}, \
  and\ \bibinfo {author} {\bibfnamefont {F.~J.}\ \bibnamefont {Cherne}},\
  }\href {http://link.aps.org/doi/10.1103/PhysRevB.66.104107} {\bibfield
  {journal} {\bibinfo  {journal} {Phys. Rev. B}\ }\textbf {\bibinfo {volume}
  {66}},\ \bibinfo {pages} {104107} (\bibinfo {year} {2002})}\BibitemShut
  {NoStop}%
\bibitem [{\citenamefont {Plimpton}(1995)}]{Plimpton1995}%
  \BibitemOpen
  \bibfield  {author} {\bibinfo {author} {\bibfnamefont {S.}~\bibnamefont
  {Plimpton}},\ }\href {\doibase 10.1006/jcph.1995.1039} {\bibfield  {journal}
  {\bibinfo  {journal} {J. Comput. Phys.}\ }\textbf {\bibinfo {volume} {117}},\
  \bibinfo {pages} {1 } (\bibinfo {year} {1995})}\BibitemShut {NoStop}%
\bibitem [{\citenamefont {Angell}\ \emph {et~al.}(1996)\citenamefont {Angell},
  \citenamefont {Borick},\ and\ \citenamefont {Grabow}}]{Angell1996}%
  \BibitemOpen
  \bibfield  {author} {\bibinfo {author} {\bibfnamefont {C.~A.}\ \bibnamefont
  {Angell}}, \bibinfo {author} {\bibfnamefont {S.}~\bibnamefont {Borick}}, \
  and\ \bibinfo {author} {\bibfnamefont {M.}~\bibnamefont {Grabow}},\
  }\href@noop {} {\bibfield  {journal} {\bibinfo  {journal} {J. Non-Cryst.
  Solids}\ }\textbf {\bibinfo {volume} {207}},\ \bibinfo {pages} {463}
  (\bibinfo {year} {1996})}\BibitemShut {NoStop}%
\bibitem [{\citenamefont {Jakse}\ \emph {et~al.}(2003)\citenamefont {Jakse},
  \citenamefont {Hennet}, \citenamefont {Price}, \citenamefont {Krishnan},
  \citenamefont {Key}, \citenamefont {Artacho}, \citenamefont {Glorieux},
  \citenamefont {Pasturel},\ and\ \citenamefont {Saboungi}}]{Jakse2003}%
  \BibitemOpen
  \bibfield  {author} {\bibinfo {author} {\bibfnamefont {N.}~\bibnamefont
  {Jakse}}, \bibinfo {author} {\bibfnamefont {L.}~\bibnamefont {Hennet}},
  \bibinfo {author} {\bibfnamefont {D.~L.}\ \bibnamefont {Price}}, \bibinfo
  {author} {\bibfnamefont {S.}~\bibnamefont {Krishnan}}, \bibinfo {author}
  {\bibfnamefont {T.}~\bibnamefont {Key}}, \bibinfo {author} {\bibfnamefont
  {E.}~\bibnamefont {Artacho}}, \bibinfo {author} {\bibfnamefont
  {B.}~\bibnamefont {Glorieux}}, \bibinfo {author} {\bibfnamefont
  {A.}~\bibnamefont {Pasturel}}, \ and\ \bibinfo {author} {\bibfnamefont
  {M.~L.}\ \bibnamefont {Saboungi}},\ }\href {\doibase 10.1063/1.1631388}
  {\bibfield  {journal} {\bibinfo  {journal} {Appl. Phys. Lett.}\ }\textbf
  {\bibinfo {volume} {83}},\ \bibinfo {pages} {4734} (\bibinfo {year}
  {2003})}\BibitemShut {NoStop}%
\bibitem [{\citenamefont {Vasisht}\ \emph {et~al.}(2011)\citenamefont
  {Vasisht}, \citenamefont {Saw},\ and\ \citenamefont {Sastry}}]{Vasisht2011}%
  \BibitemOpen
  \bibfield  {author} {\bibinfo {author} {\bibfnamefont {V.~V.}\ \bibnamefont
  {Vasisht}}, \bibinfo {author} {\bibfnamefont {S.}~\bibnamefont {Saw}}, \ and\
  \bibinfo {author} {\bibfnamefont {S.}~\bibnamefont {Sastry}},\ }\href
  {http://dx.doi.org/10.1038/nphys1993} {\bibfield  {journal} {\bibinfo
  {journal} {Nat. Phys.}\ }\textbf {\bibinfo {volume} {7}},\ \bibinfo {pages}
  {549} (\bibinfo {year} {2011})}\BibitemShut {NoStop}%
\bibitem [{\citenamefont {de~Koning}\ \emph {et~al.}(2009)\citenamefont
  {de~Koning}, \citenamefont {Antonelli},\ and\ \citenamefont
  {Jara}}]{deKoning2009}%
  \BibitemOpen
  \bibfield  {author} {\bibinfo {author} {\bibfnamefont {M.}~\bibnamefont
  {de~Koning}}, \bibinfo {author} {\bibfnamefont {A.}~\bibnamefont
  {Antonelli}}, \ and\ \bibinfo {author} {\bibfnamefont {D.~A.~C.}\
  \bibnamefont {Jara}},\ }\href
  {http://link.aps.org/doi/10.1103/PhysRevB.80.045209} {\bibfield  {journal}
  {\bibinfo  {journal} {Phys. Rev. B}\ }\textbf {\bibinfo {volume} {80}},\
  \bibinfo {pages} {045209} (\bibinfo {year} {2009})}\BibitemShut {NoStop}%
\bibitem [{\citenamefont {Hansen}\ and\ \citenamefont
  {McDonald}(2006)}]{Hansen2006}%
  \BibitemOpen
  \bibfield  {author} {\bibinfo {author} {\bibfnamefont {J.}~\bibnamefont
  {Hansen}}\ and\ \bibinfo {author} {\bibfnamefont {I.}~\bibnamefont
  {McDonald}},\ }\href {http://books.google.com.br/books?id=Uhm87WZBnxEC}
  {\emph {\bibinfo {title} {Theory of Simple Liquids}}}\ (\bibinfo  {publisher}
  {Elsevier Science},\ \bibinfo {year} {2006})\BibitemShut {NoStop}%
\bibitem [{\citenamefont {Daisenberger}\ \emph {et~al.}(2011)\citenamefont
  {Daisenberger}, \citenamefont {Deschamps}, \citenamefont {Champagnon},
  \citenamefont {Mezouar}, \citenamefont {Quesada~Cabrera}, \citenamefont
  {Wilson},\ and\ \citenamefont {McMillan}}]{Daisenberger2011}%
  \BibitemOpen
  \bibfield  {author} {\bibinfo {author} {\bibfnamefont {D.}~\bibnamefont
  {Daisenberger}}, \bibinfo {author} {\bibfnamefont {T.}~\bibnamefont
  {Deschamps}}, \bibinfo {author} {\bibfnamefont {B.}~\bibnamefont
  {Champagnon}}, \bibinfo {author} {\bibfnamefont {M.}~\bibnamefont {Mezouar}},
  \bibinfo {author} {\bibfnamefont {R.}~\bibnamefont {Quesada~Cabrera}},
  \bibinfo {author} {\bibfnamefont {M.}~\bibnamefont {Wilson}}, \ and\ \bibinfo
  {author} {\bibfnamefont {P.~F.}\ \bibnamefont {McMillan}},\ }\href {\doibase
  10.1021/jp205090s} {\bibfield  {journal} {\bibinfo  {journal} {J. Phys. Chem.
  B}\ }\textbf {\bibinfo {volume} {115}},\ \bibinfo {pages} {14246} (\bibinfo
  {year} {2011})}\BibitemShut {NoStop}%
\bibitem [{\citenamefont {Allen}\ and\ \citenamefont
  {Tildesley}(1989)}]{Allen1989}%
  \BibitemOpen
  \bibfield  {author} {\bibinfo {author} {\bibfnamefont {M.}~\bibnamefont
  {Allen}}\ and\ \bibinfo {author} {\bibfnamefont {D.}~\bibnamefont
  {Tildesley}},\ }\href {http://books.google.com.br/books?id=O32VXB9e5P4C}
  {\emph {\bibinfo {title} {Computer simulation of liquids}}},\ Oxford science
  publications\ (\bibinfo  {publisher} {Clarendon Press},\ \bibinfo {year}
  {1989})\BibitemShut {NoStop}%
\bibitem [{\citenamefont {Alf\'{e}}\ and\ \citenamefont
  {Gillan}(1998)}]{Alfe1998}%
  \BibitemOpen
  \bibfield  {author} {\bibinfo {author} {\bibfnamefont {D.}~\bibnamefont
  {Alf\'{e}}}\ and\ \bibinfo {author} {\bibfnamefont {M.~J.}\ \bibnamefont
  {Gillan}},\ }\href {http://link.aps.org/doi/10.1103/PhysRevLett.81.5161}
  {\bibfield  {journal} {\bibinfo  {journal} {Phys. Rev. Lett.}\ }\textbf
  {\bibinfo {volume} {81}},\ \bibinfo {pages} {5161} (\bibinfo {year}
  {1998})}\BibitemShut {NoStop}%
\bibitem [{\citenamefont {Rhim}\ and\ \citenamefont {Ohsaka}(2000)}]{Rhim2000}%
  \BibitemOpen
  \bibfield  {author} {\bibinfo {author} {\bibfnamefont {W.-K.}\ \bibnamefont
  {Rhim}}\ and\ \bibinfo {author} {\bibfnamefont {K.}~\bibnamefont {Ohsaka}},\
  }\href {\doibase 10.1016/S0022-0248(99)00437-6} {\bibfield  {journal}
  {\bibinfo  {journal} {J. Cryst. Growth}\ }\textbf {\bibinfo {volume} {208}},\
  \bibinfo {pages} {313 } (\bibinfo {year} {2000})}\BibitemShut {NoStop}%
\bibitem [{\citenamefont {G\"{o}tze}(1991)}]{Hansen1991}%
  \BibitemOpen
  \bibfield  {author} {\bibinfo {author} {\bibfnamefont {W.}~\bibnamefont
  {G\"{o}tze}},\ }\href {http://books.google.com.br/books?id=5AcvAQAAIAAJ}
  {\emph {\bibinfo {title} {Liquids, Freezing and Glass Transition, Les
  Houches, Session LI, 1989}}},\ edited by\ \bibinfo {editor} {\bibfnamefont
  {J.}~\bibnamefont {Hansen}}, \bibinfo {editor} {\bibfnamefont
  {D.}~\bibnamefont {Levesque}}, \ and\ \bibinfo {editor} {\bibfnamefont
  {J.}~\bibnamefont {Zinn-Justin}},\ \bibinfo {number} {v. 1-2}\ (\bibinfo
  {publisher} {North Holland},\ \bibinfo {year} {1991})\ p.\ \bibinfo {pages}
  {287}\BibitemShut {NoStop}%
\bibitem [{\citenamefont {Vogel}(1921)}]{Vogel1921}%
  \BibitemOpen
  \bibfield  {author} {\bibinfo {author} {\bibfnamefont {H.}~\bibnamefont
  {Vogel}},\ }\href@noop {} {\bibfield  {journal} {\bibinfo  {journal} {Phys.
  Z.}\ }\textbf {\bibinfo {volume} {22}},\ \bibinfo {pages} {645} (\bibinfo
  {year} {1921})}\BibitemShut {NoStop}%
\bibitem [{\citenamefont {Fulcher}(1925)}]{Fulcher1925}%
  \BibitemOpen
  \bibfield  {author} {\bibinfo {author} {\bibfnamefont {G.~S.}\ \bibnamefont
  {Fulcher}},\ }\href {\doibase 10.1111/j.1151-2916.1925.tb16731.x} {\bibfield
  {journal} {\bibinfo  {journal} {J. Amer. Ceram. Soc.}\ }\textbf {\bibinfo
  {volume} {8}},\ \bibinfo {pages} {339} (\bibinfo {year} {1925})}\BibitemShut
  {NoStop}%
\bibitem [{\citenamefont {Tammann}\ and\ \citenamefont
  {Hesse}(1926)}]{Tammann1926}%
  \BibitemOpen
  \bibfield  {author} {\bibinfo {author} {\bibfnamefont {G.}~\bibnamefont
  {Tammann}}\ and\ \bibinfo {author} {\bibfnamefont {W.}~\bibnamefont
  {Hesse}},\ }\href@noop {} {\bibfield  {journal} {\bibinfo  {journal} {Z.
  Anorgan. Allg. Chem.}\ }\textbf {\bibinfo {volume} {156}} (\bibinfo {year}
  {1926})}\BibitemShut {NoStop}%
\bibitem [{\citenamefont {Spells}(1936)}]{Spells1936}%
  \BibitemOpen
  \bibfield  {author} {\bibinfo {author} {\bibfnamefont {K.~E.}\ \bibnamefont
  {Spells}},\ }\href {http://stacks.iop.org/0959-5309/48/i=2/a=308} {\bibfield
  {journal} {\bibinfo  {journal} {P. Phys. Soc.}\ }\textbf {\bibinfo {volume}
  {48}},\ \bibinfo {pages} {299} (\bibinfo {year} {1936})}\BibitemShut
  {NoStop}%
\bibitem [{\citenamefont {Angell}(1997)}]{Angell1997}%
  \BibitemOpen
  \bibfield  {author} {\bibinfo {author} {\bibfnamefont {C.~A.}\ \bibnamefont
  {Angell}},\ }\href {\doibase 10.6028/jres.102.013} {\bibfield  {journal}
  {\bibinfo  {journal} {J. Res. NIST}\ }\textbf {\bibinfo {volume} {102}},\
  \bibinfo {pages} {171} (\bibinfo {year} {1997})}\BibitemShut {NoStop}%
\end{thebibliography}

%

\end{document}